\documentclass[10pt,twocolumn]{paper}
\usepackage{epsfig, graphicx}
\usepackage[english]{babel}
\usepackage[margin=1.5cm]{geometry}

\title{\center \rm \bf Synchrotron X-Ray Reflectivity Study of the Adsorption Film of Octadecanamide 
at the Toluene -- Water Interface}

\author{\small \rm Aleksey M. Tikhonov$^{a,b}$\/\thanks{tikhonov@kapitza.ras.ru} {} and Yurii O. Volkov$^{b,c}$\/\thanks{neko.crys@gmail.com}\\
\small $^a$Kapitza Institute for Physical Problems, Russian Academy of Sciences, Moscow, Russia\\
\small $^b$Institute of Solid State Physics, Russian Academy of Sciences, Chernogolovka, Moscow region, 142432 Russia\\
\small $^c$Shubnikov Institute of Crystallography, Federal Research Center Crystallography and Photonics,
Russian Academy of Sciences, \\
\small Moscow, 119333 Russia\\
}

\begin{document}

\maketitle

\abstract{ \it  The structure of an adsorption octadecanamide film at the planar toluene -- water interface is studied
by X-ray reflectometry using synchrotron radiation with photon energy of 15 keV. The electron density
(polarizability) profiles, according to which the interface structure is determined by the pH level in the water
subphase, are reconstructed from experimental data with the help of a model-independent approach. For a
high pH$\approx 11$, the adsorption film is a crystalline octadecanamide monolayer with a thickness of about 26~{\AA},
in which aliphatic tails of surfactant are extended along the normal to the surface. For low pH$\approx 2$, the thickness
of the surface structure consisting of the crystalline monolayer directly on the toluene -- water interface
and a thick layer of deposited octadecanamide micelles reaches about 500~{\AA}. In our opinion, the condensation
of nonionogenic surfactant micelles for which the surface concentration of the surfactant increases significantly
is caused by a change in the polarization direction upon a decrease in the pH level in the electric
double layer at the interface between the water subphase and the octadecanamide monolayer. The shape of
the reconstructed electron density profiles also indicates the existence of a plane of the closest approach of
surfactant micelles to the interface at a distance of about 70~{\AA} from it.  }

\vspace{0.25in}
\normalsize

{\bf INTRODUCTION}

The experimental determination of the structure of the transition layer at the interface between two condensed
phases (in particular, a nonpolar organic solvent (oil) and water) is an important problem in the
physics of surface phenomena. A soluble adsorption film of a diphyllic substance (surfactant) on this surface
can be treated as a 2D thermodynamic system in which, for example, various barotropic, lyatropic, and
thermotropic phase transitions between surface mesophases can be observed \cite{1,2,Alconols0,MingLi3,ScHigh,MingLi2,Alconols,Mixed,SCHTAM,3,crossover,S-L,Smectic}. 
The properties of epitropic
liquid-crystal layers with a similar structure determine viscosity parameters of motor oils in the tribotriad
\cite{4, 5}. The application of synchrotron radiation in the hard part of the X-ray wavelength range in
analysis of the structure of buried interfaces using the reflectometry and diffuse scattering methods provides
basically new experimental tools for determining the nature of these and many other phenomena in surface
layers \cite{MRS,6,alco-c30,7,Tail,8,HS,WD,9,azacrown,10,11,12,13}.

\begin{figure}
\hspace{0.15in}
\epsfig{file=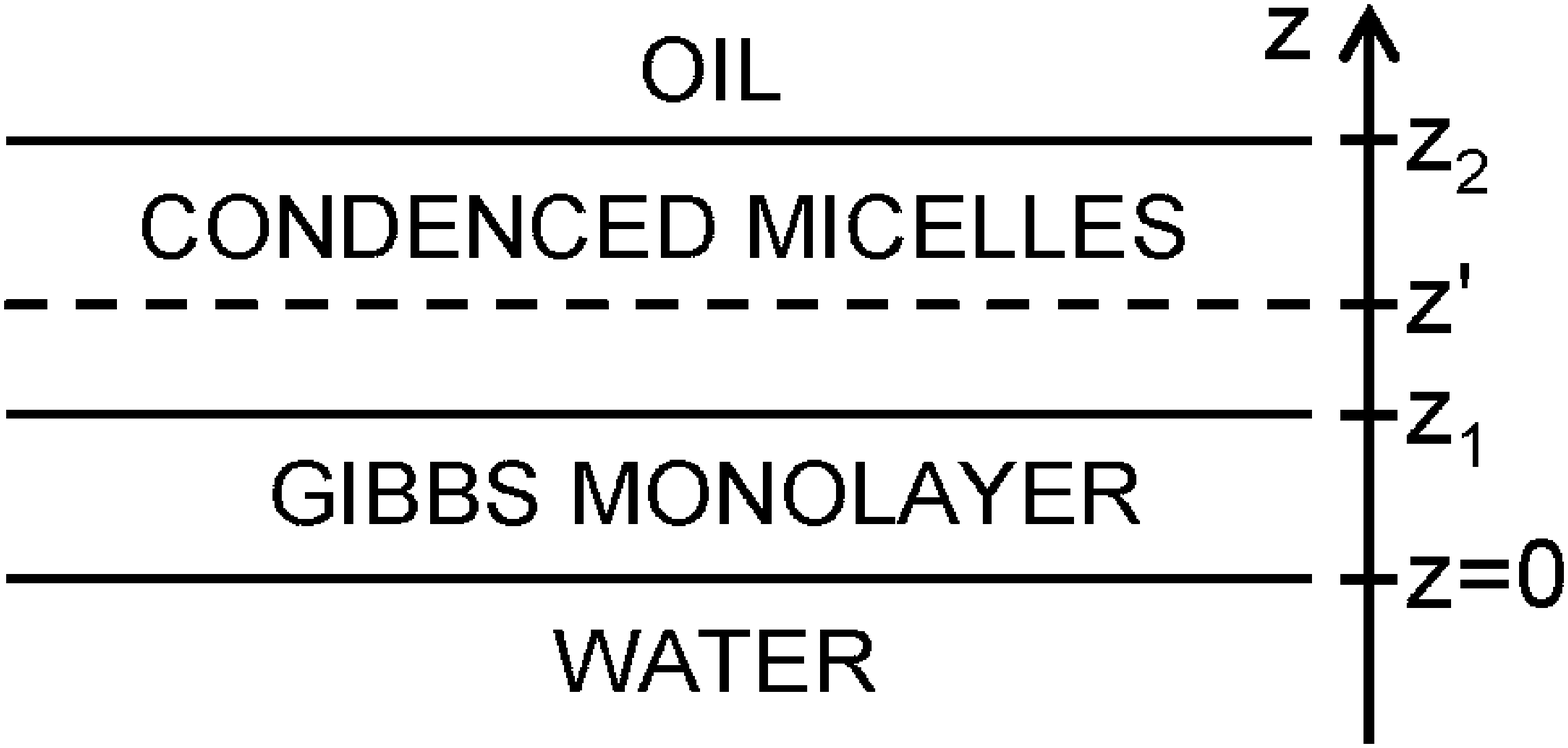, width=0.45\textwidth}

\small {\bf Figure 1.} \it Structure of the adsorption film on the oil -- water interface; $z_1$ is the thickness of the Gibbs monolayer, $z\prime$ is the position of the plane of closest approach of micelles to the interface; and $z_2$ is the diffuse boundary between the
condensed micelle layer and the bulk nonpolar solvent.

\end{figure}

Earlier, there were reports on the observation of transitions to multilayer adsorption in two-component
adsorption fluoroalkanol films and in one-component triacontanol and triacontanic acid films at the
n-hexane -- water interface (see, for instance, \cite{14}). In our recent publication \cite{13,15}, it was shown that
upon a decrease in temperature $T$, the 2D condensation transition, for example, of triacontanol from the
gas phase to the liquid Gibbs monolayer of thickness $z_1 \approx  27$~{\AA} at the n-hexane -- water interface is followed
by its transition to the monolayer adsorption, which we attributed to the increase in the micelle concentration
in the surface layer of thickness $z_2 \approx 200$~{\AA} \cite{16} (Fig. 1). The structural information is obtained from
the results of these and other experiments based on the calculation of the X-ray reflectivity for model surface
structures using available information, for example, on the geometrical sizes and structure of molecules,
interfacial tension, etc.

In this study, we analyze experimental data for the interface using a model-independent approach that
does not require any additional assumptions concerning the transverse structure of the surface \cite{17,IVK2000,IVK2012a,condensation}. The X-ray reflectometry data are used for studying the structure of the octadecanamide adsorption film on
the planar toluene -- water interface depending on the composition of the water subphase (pH level in it). It is found that at a low pH level (not exceeding 7), the surface concentration and thickness of the adsorption film of the amphiphylic substance increase significantly
(by several times), which we attribute to the condensation of its micelles, induced by their electrostatic
interaction with the interface. The application of the new approach has made it possible, for example, to
establish the existence and determine the position of the plane of the closest approach of micelles to the
interface (see Fig. 1).

\vspace{0.25in}
{\bf EXPERIMENTAL}

We have earlier established experimentally that the minimal admissible thickness (along the beam) of the
hydrocarbon phase in the experimental cell in analysis of the oil -- water interface is about 75 mm \cite{18,HighPP} (Fig. 2).
For the normally incident photons (for example, with energy of 8 keV), the coefficient of transmission
through such a hydrocarbon layer is about 10$^{-11}$. In a direct beam with an input intensity of about 10$^7$~photon/s
(wide-focus X-ray tube with a copper anode), we obtain 10$^{-3}$~photon/s at the sample exit, which is
2 - 3 orders of magnitude lower than the intrinsic noise level of a modern X-ray detector. For harder (e.g.,
15 keV) photons, the transmission coefficient  increases to approximately 10$^{-2}$. However, a probe
beam with a vertical width of smaller than 10~$\mu$m and an angular divergence of 10$^{-5}$~rad is used in actual
practice for correct measurement of the reflectivity factor from the interface. The preparation of a beam
with such parameters inevitably leads to a considerable decrease in intensity. For this reason, laboratory
sources for X-ray diffraction studies of buried planar interfaces are less attractive considering their low
brightness as compared to synchrotron sources of hard radiation \cite{19}.

\begin{figure}
\hspace{0.15in}
\epsfig{file=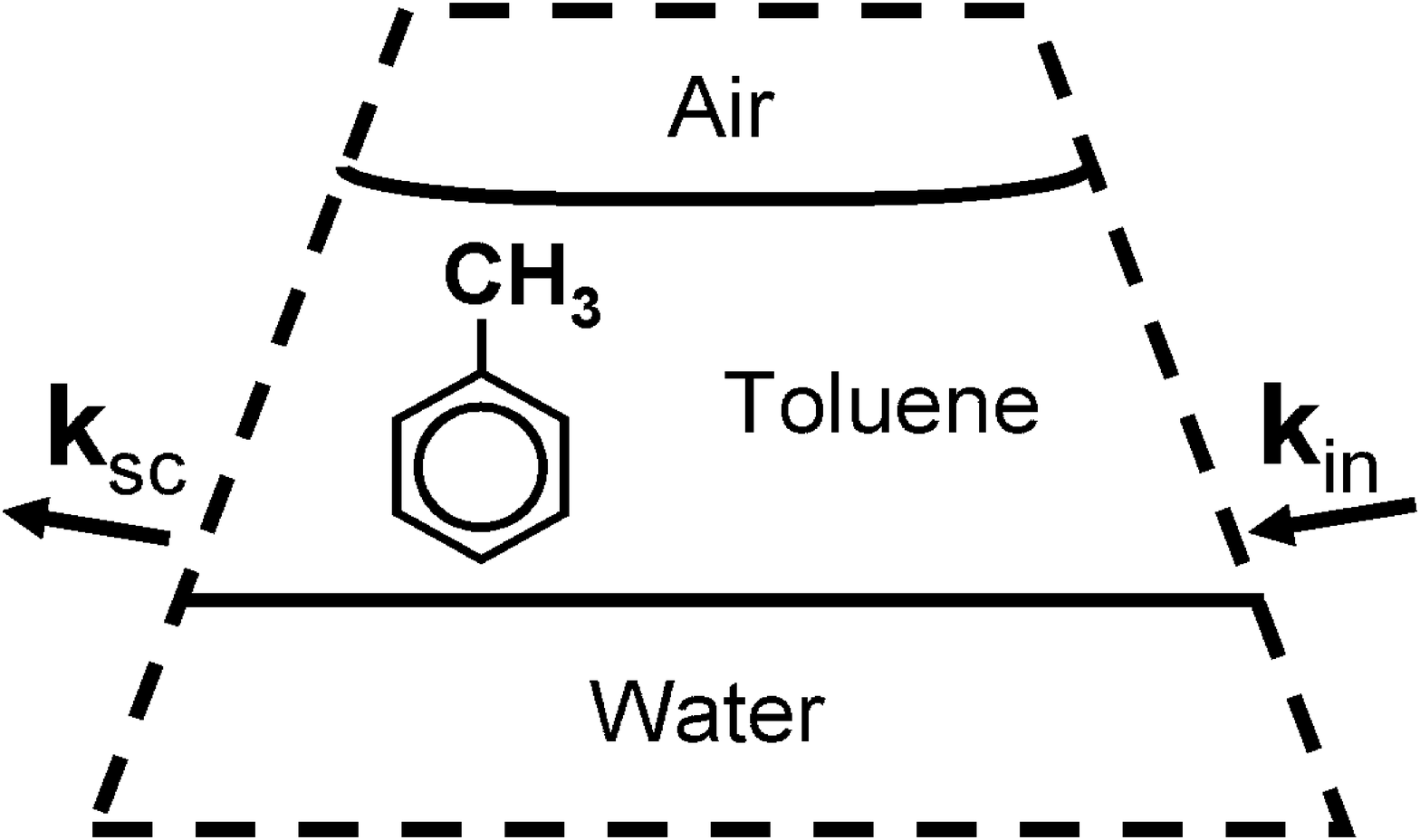, width=0.4\textwidth}

\small {\bf Figure 2.} \it Schematic diagram of the experimental cell.

\end{figure}

The samples with a planar toluene -- water interface oriented by the gravity force were studied under standard conditions in a thermocycled stainless steel cell in accordance with the technique described in \cite{20,t-w3}. X-ray-transparent windows of the cell were prepared from polyester; the geometrical sizes of the interface between bulk phases in such a cell are 75~mm~$\times$~150~mm along and across the beam, respectively.

All chemical components for experiments were purchased at the Sigma-Aldrich company. We used octadecanamide C$_{18}$H$_{37}$NO as a nonionogenic surfactant that is well-soluble in the aromatic hydrocarbon
and is insoluble in water. The estimated length of this linear chain molecule is $L \approx 26$~{\AA} ($= 17 \times 1.27$~{\AA}
(C-C) + 1.5 ~{\AA} (-CH3) + 2.5~{\AA} (-CONH$_2$))~\cite{21}.

Toluene (C$_7$H$_8$ with a density of approximately 0.86~g/cm$^3$ at $T = 298$~K; boiling point is $T_b = 384$~K) was purified in a chromatographic column \cite{22}. The upper oil phase was a solution of octadecanamide in toluene about 75 ml in volume and with a volume concentration of about 5 mmol/L. Deionized water with pH$\approx 7$ (Barnstead, NanoPureUV), sulfuric acid solution
(pH$\approx 2$), and NaOH solution (pH$\approx 11$) in it of about 100~ml in volume were used as the lower bulk
water phase. Prior to experiments, liquids were degassed in an ultrasonic bath. During measurements of X-ray reflectivity $R$, the sample was subjected to "annealing": the temperature of liquids in the cell was increased by 30~K above the room temperature and
then the sample was brought to equilibrium at $T = 298$\,K by thorough mechanical stirring for several hours \cite{13}.

\begin{figure}
\hspace{0.15in}
\epsfig{file=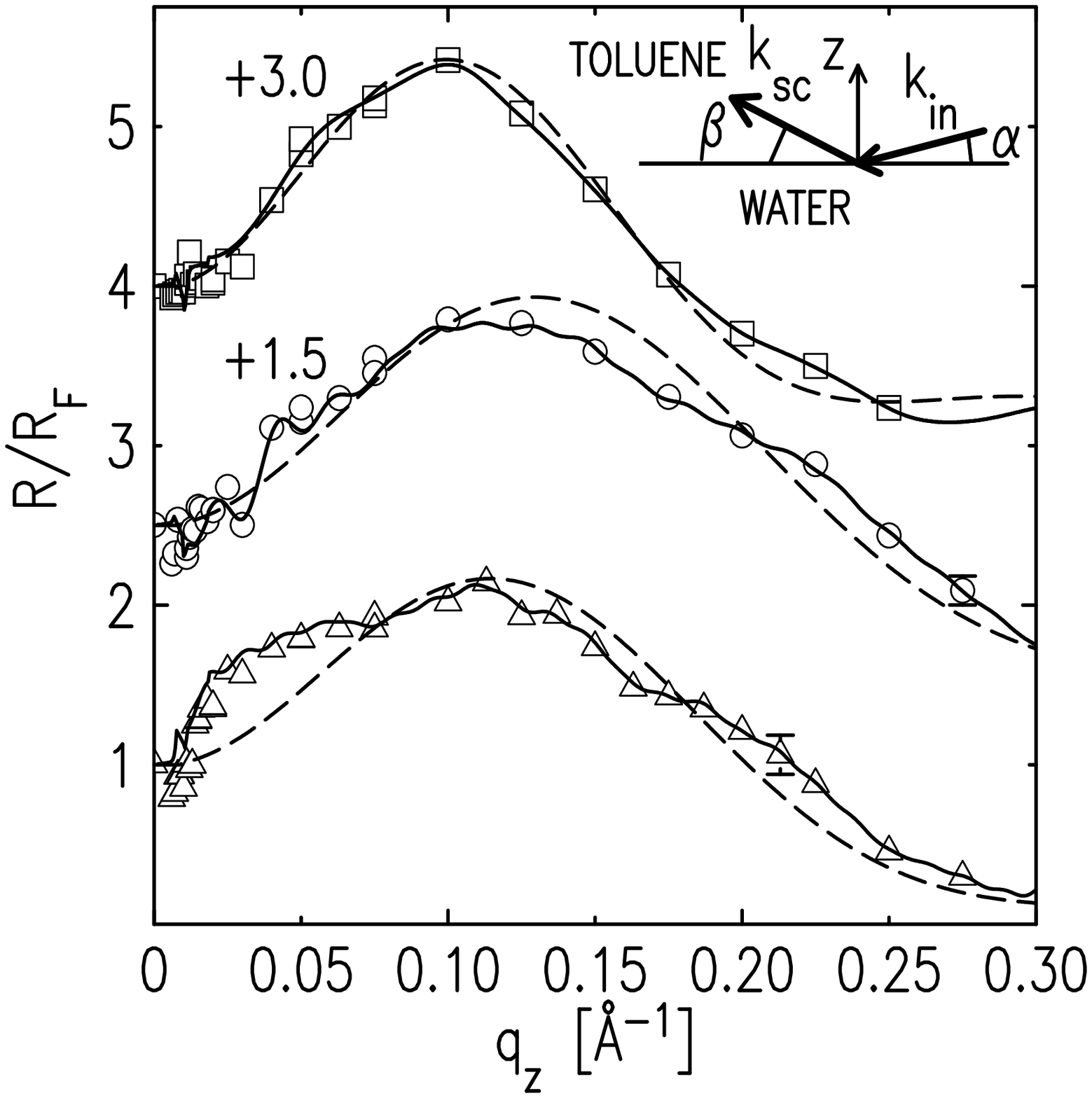, width=0.45\textwidth}

\small {\bf Figure 3.} \it Normalized reflectivity factor $R/R_F$ as a function of $q_z$ for the interfaces with the adsorption octadecanamide film; triangles, circles, and squares show the boundary for pH $\approx 2$, 7, and 11, respectively. Solid and dashed curves describe the results of model-independent and model calculations, respectively. Numbers on the curves indicate
their displacement along the ordinate axis for better visualization of representation of results. The inset shows the
kinematics of X-ray surface scattering at the toluene -- water interface. Reflectivity is measured at $\alpha=\beta$.

\end{figure}

Interfacial tension $\gamma$ was measured by the Wilhelmy plate method directly in the experimental cell \cite{23}. For this purpose, we used a chromatographic paper (Whatman) sheet of size approximately  $\approx 10\times5$\,mm which was fixed to a platinum wire (of about $0.25$~mm in diameter) passing through the holes (of diameter about 2~mm) in the upper lid of the thermostat and in the cell port cover. The covers were replaced by the lids consisting of two parts, which expectedly affected the leak tightness of the cell in tension measurements. Interfacial tension $\gamma\approx\Delta F/2L$ was determined from change $\Delta F$ in the weight of the plate during its separation from the toluene -- water interface, which was measured by a NIMA PS-2 electric balance. After its
separation from the interface, the plate remains completely submerged in the hydrocarbon phase. At $T = 298$~K, the value of $\gamma$ for all toluene samples lies in interval $23 \div 24$\,mN/m, which is smaller by approximately
1/3 than a tension of $36.0 \pm 0.1$\,mN/m at the toluene -- water interface between pure liquids \cite{24, 25}.

The transverse structure of the toluene -- water interface was studied by reflectometry at the X19C station
of the NSLS synchrotron \cite{26}. In experiments, a focused monochromatic photon beam of intensity about $10^{11}$~photon/s and energy $E = 15$~keV ($\lambda=0.825 \pm 0.002$ \,\AA) was used. The structure of the station made it possible to study the surfaces of solids as well as liquids \cite{27,28,29}. For example, we have investigated earlier a transition of melting at the toluene -- water interface in the adsorbed monolayer of octodecane acid using this setup \cite{30}.

In the case of specular reflection, scattering vector has only one nonzero component $q_z=(4\pi/\lambda)\sin\alpha$ along the normal to the surface (see inset to Fig. 3). For grazing angles$\alpha$ smaller than critical value $\alpha_c\approx\lambda\sqrt{r_e\Delta\rho/\pi}$ (where $r_e = 2.814\cdot10^{-5}$\,{\AA} is the classical
electron radius and $\Delta\rho = \rho_w - \rho_{t}$), the incident beam experiences total external reflection, and $R\approx 1$. Under standard conditions, the electron density is as much as $\rho_w\approx 0.333$\,{\it e$^-$/}{\AA}$^3$ ({\it e$^-$} is the electron charge)  in water and $\rho_{t} \approx 0.85 \rho_w$ in toluene ($\Delta\rho \approx 0.15 \rho_w$). Therefore, $\alpha_c\approx 0.03$\,deg at the toluene -- water interface.

Figure 3 shows the experimental dependences of reflectivity factor $R$ on $q_z$, which are normalized for
better visualization by Fresnel function  $R_F(q_z)\approx$ $(q_z-[q_z^2-q_c^2]^{1/2})^2/(q_z+[q_z^2-q_c^2]^{1/2})^2$,
where $q_c=(4\pi/\lambda)\sin\alpha_c$$\approx 0.0085$\,\AA$^{-1}$. Triangles correspond
to values of $R(q_z)/R_F(q_z)$ at a level of pH$\approx 2$, circles correspond to pH$\approx 7$, and squares, to pH$\approx$11.
These data demonstrate a quite strong dependence of the interface structure on the pH level in the water subphase.

\vspace{0.25in}
{\bf THEORY}

According to experimental data on $R(q_z)$ in the interval from approximately 1 to $10^{-9}$, we reconstructed
electron density distributions $\rho(z)$ along the normal to the surface using two fundamentallly different
approaches, viz., with the assumption concerning the monolayer structure of the adsorption film and
without any assumption about the transverse structure of the surface. In the former case, we used a qualitative
monolayer model based on the error function, which describes temperature-activated fluctuations of the
interface (capillary waves) \cite{31,32,33,34}. In the first Born approximation of distorted waves (DWBA), reflectivity factor $R(q_z)$ for the toluene -- water interface with the Gibbs monolayer
is expressed as follows \cite{35, 36}:
\begin{equation}
\begin{array}{l}
\displaystyle \frac{R(q_z)}{R_F(q_z)} \approx \frac{\exp\left(-\sigma^2q_zq^t_z\right)}{\Delta\rho^2}
\\ \\
\displaystyle
\times\left|\rho_1 - \rho_w + (\rho_t-\rho_1)\exp\left(iz_1\sqrt{q_zq^t_z}\right)\right|^2,
\end{array}
\end{equation}

where $q^t_z=\sqrt{q_z^2-q_c^2}$ and $\rho_1$ is the electron density in the octadecanamide monolayer. The position of the monolayer -- water interface corresponds to $z_0 = 0$, $z_1$ is the thickness of the Gibbs monolayer (see Fig. 1), and
$\sigma$ is the standard deviation of the positions of the boundaries from their nominal values $z_0$ and $z_1$. In our
calculations, we fixed the value of parameter $\sigma^2$ for the monolayer boundaries equal to the square of the "capillary
width": $\sigma^2 = (k_BT/2\pi\gamma)\ln(Q_{max}/Q_{min})$, where $k_B$ is the Boltzmann constant, the short-wavelength
limit in the capillary wave spectrum is $Q_{max} = 2\pi/a$ ($a\approx 10$\,\AA{} is of the same order of magnitude as the
intermolecular spacing), and the long-wavelength limit of fluctuations of the surface in the experiment is
$Q_{min}=q_z^{max}\Delta\beta$ (where $q_z^{max}\approx 0.275$\,\AA$^{-1}$ and the detector
angular resolution is $\Delta\beta$$\approx 4\cdot10^{-4}$\,rad) \cite{37}. In our experimental conditions, the calculated value of the
width for the toluene -- water interface is $\sigma = 4.75 \pm 0.05$\,{\AA}. This approach to analysis of reflectometry data
have been successfully used in our earlier investigation of structures and phase transitions in adsorption films of amphiphilic substances at  the planar air -- water interface, as well as at the  interface between saturated hydrocarbon n-hexane and water \cite{Bitto, multilayer,PL,38}.

The results of calculations for the model monolayer based on expression (1) are shown by dashed curves in
Fig. 3. For pH$\approx 11$, the electron density in the Gibbs monolayer at the toluene -- water interface is $\rho_1 = 0.350 \pm 0.005$\,{\it e$^-$/}{\AA}$^3$ and $z_1=26 \pm 1$\,{\AA}. Monolayer density $\rho_1$ corresponds to the closest packing of hydrocarbon chains in the crystalline  "$\gamma$-phase", and the monolayer thickness coincides (to within the error) with length $L$ of the octadecanamide molecule \cite{39}. Such values of fitting parameters correspond to the area per molecule in the monolayer $A=\Gamma/(z_1\rho_1)$$=18\pm1$\,\AA$^2$, where $\Gamma=160$ is the number of electrons in C$_{18}$H$_{37}$NO molecule. Therefore, we can conventionally refer to the monolayer as a crystalline layer in which aliphatic tails of the surfactant are extended along the normal to the surface.

For other pH levels, the value of fitting parameter $\rho_1$ is approximately the same as for pH $\approx 11$, but $z_1$ is
noticeably smaller than $L$ ($z_1 = 17 \div 20$\,{\AA}). On the one hand, this could correspond to the solid hexatic phase
of the monolayer with a large angle of inclination of aliphatic tails (their angle of deviation from the normal
is $\theta = \arccos(z_1/L)\approx 40^\circ$) analogously to the phases of the Langmuir monolayer of octodecane acid on the
water surface \cite{40, 41}. On the other hand, the integral characteristic of the monolayer is the area per molecule
$A=\Gamma/(z_1\rho_1)$$\approx 26$\,\AA$^2$, which corresponds to liquid octodecane (C$_{18}$H$_{38}$). This obvious contradiction is undoubtedly because of a more complex structure of the adsorption film than a monolayer.

The fact that the square of structure factor $R/R_F$ exceeds unity for small gliding angles (see data for
pH $\approx 2$) in the interval of $q_z$ values from approximately $1.5q_c$ to $5q_c$ indicates the presence of the region of
excessive electron density of the adsorbed substance near the interface as compared to the density of the
model monolayer, which we attribute to the condensation of octadecanamide micelles at the interface. To
relate this peculiarity on the $R(q_z)$ curves to the structure of the toluene -- water interface, we applied the
model-independent approach based on the extrapolation of the asymptotic behavior of reflectivity factor
$R(q_z)$ to the region of large values of $qz$ \cite{42}. We successfully used this approach earlier in describing the
structure of phospholipid multiplayers on the colloidal silica sol \cite{43, 44,IS3,Volkov} and lipid monolayers on a water
substrate \cite{DMPS1,45,Langmuir}.

\begin{figure}
\hspace{0.15in}
\epsfig{file=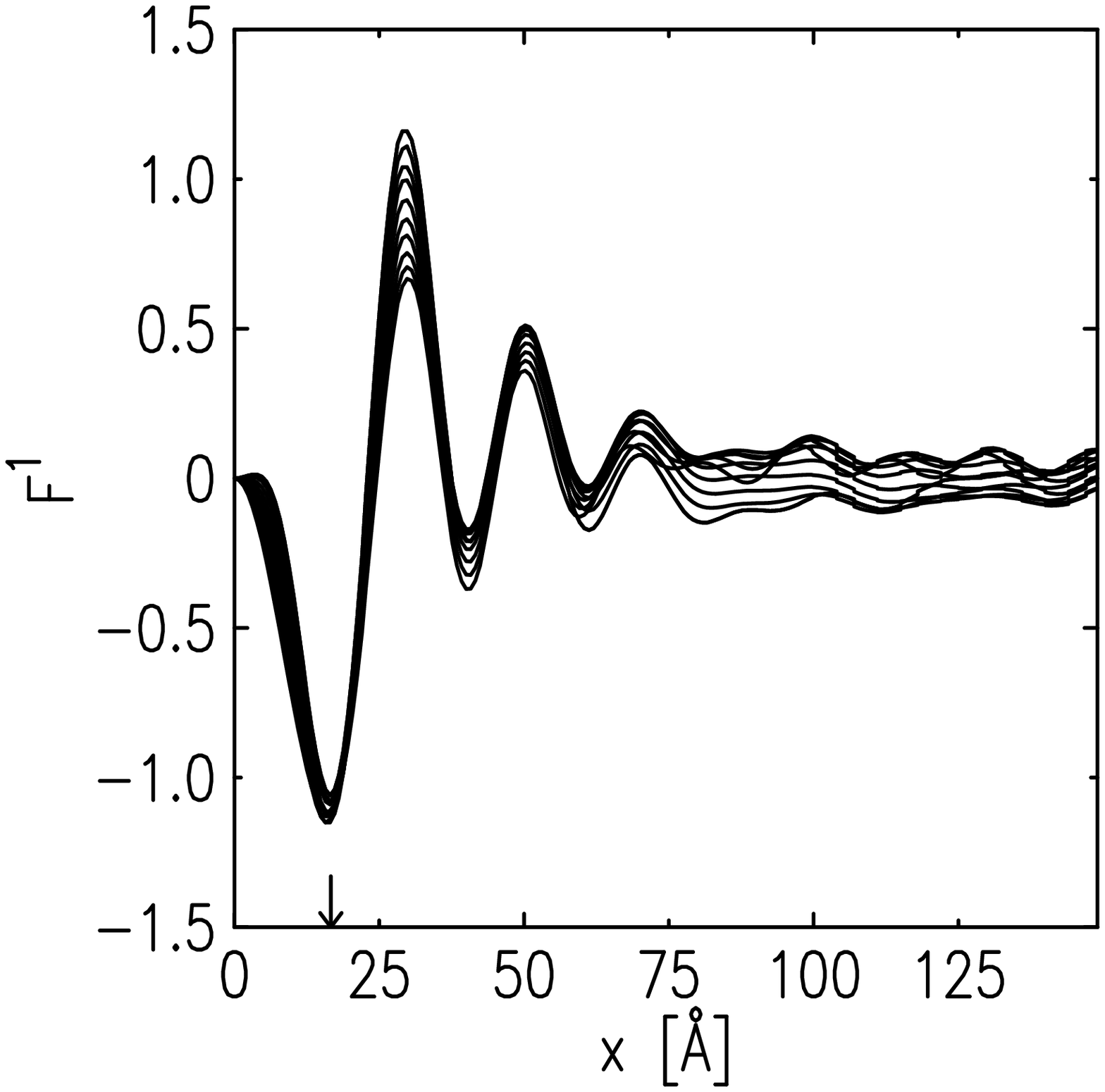, width=0.45\textwidth}

\small {\bf Figure 4.} \it Calculated functions $F^{1}(x)$ for different values of $q\prime$ and $q^\prime\prime$ for the reflectivity curve from the octadecanamide film at the toluene -- water interface for pH~$\approx 11$. The arrow indicates the position of the stable minimum at $x = 15.6$~{\AA}, which corresponds to a pair of singular points.

\end{figure}

The key advantage of this approach is that it does not require any a priori assumptions concerning the
structure of the object under investigation and provides  the absolute depth-wise distribution of polarizability
of the medium $\delta(z)$ (the real part of the complex permittivity) and, accordingly, electron density
$\rho(z)\simeq\pi\delta(z)/(r_e\lambda)$. In the general case, distribution $\delta(z)$ is represented as a piecewise-continuous function containing singular points $\Delta^{(n)}(z_j)$ at which its $n$-th derivative varies stepwise; the asymptotic form of
the decrease of reflection curve in this case is $R(q_z\to\infty)\propto 1/q_z^{(2n+4)}$.

The procedure of further analysis was described in detail in \cite{17, 42}. It should only be noted that all experimental
curves in our case decrease in proportion to $\propto 1/q_z^6$, indicating the existence of singular points of the first order ($n=1$). The mutual positions of the curves can be determined using the following procedure of modified Fourier analysis:
\begin{equation}
  \begin{array}{l}
  \displaystyle
    F^1(x) = \frac{64}{k^4(q^{\prime\prime}-q^\prime)}
    \\ \\
   \displaystyle
    \times\int\limits_{q^\prime}^{q^{\prime\prime}}\left[ q^{6}R(q) - C \right]\cos(2qx)dq;
    \\ \\
     \displaystyle
    C = \frac{1}{q^{\prime\prime}-q^\prime}\int\limits_{q^\prime}^{q^{\prime\prime}}q^{6}R(q)dq,
  \end{array}
\end{equation}
where $k=2\pi/\lambda$. The integration in Eq. 2 with respect to $q \equiv q_z/2$ is performed for different values of $q^\prime$ and $q^{\prime\prime}$. Stable extrema of function $F^1(x)$ correspond to paired distances between singular points $x_{ij}$ at which $F^1(x_{ij})\sim\Delta^1_i\Delta^1_j$.

For example, Fig. 4 shows the set of $F^1(x)$ calculated from the reflection curve for pH$\approx 11$. The only
stable minimum indicates the only pair of singular points opposite in sign and separated by distance $x=15.6$\,{\AA}, which correspond to the toluene.Gibbs monolayer interface ($z_1$ in Fig. 1) and to the maximal  electron density in the vicinity of polar groups -CONH$_2$ of the octadecanamide monolayer. Further, the $\delta(z)$ profile defined numerically by set $M\sim 100$ of thin homogeneous layers with fixed positions of singular points $z_j$ was reconstructed by fitting of the calculated angular dependence of reflectivity factor
$R_{calc}(q_z,\delta(z))$ to experimental data $R(q_z)$ using the standard least-square minimization method. To ensure
stability of the solution, the target function was supplemented with a regularization term of form $\sum\limits_{m\neq j}^M(\delta_{m-1}-\delta_{m})\to\rm{min}$, which determines the smoothness of profile $\delta(z_1\ldots z_M)$ in the intervals between singular points.

\begin{figure}
\hspace{0.15in}
\epsfig{file=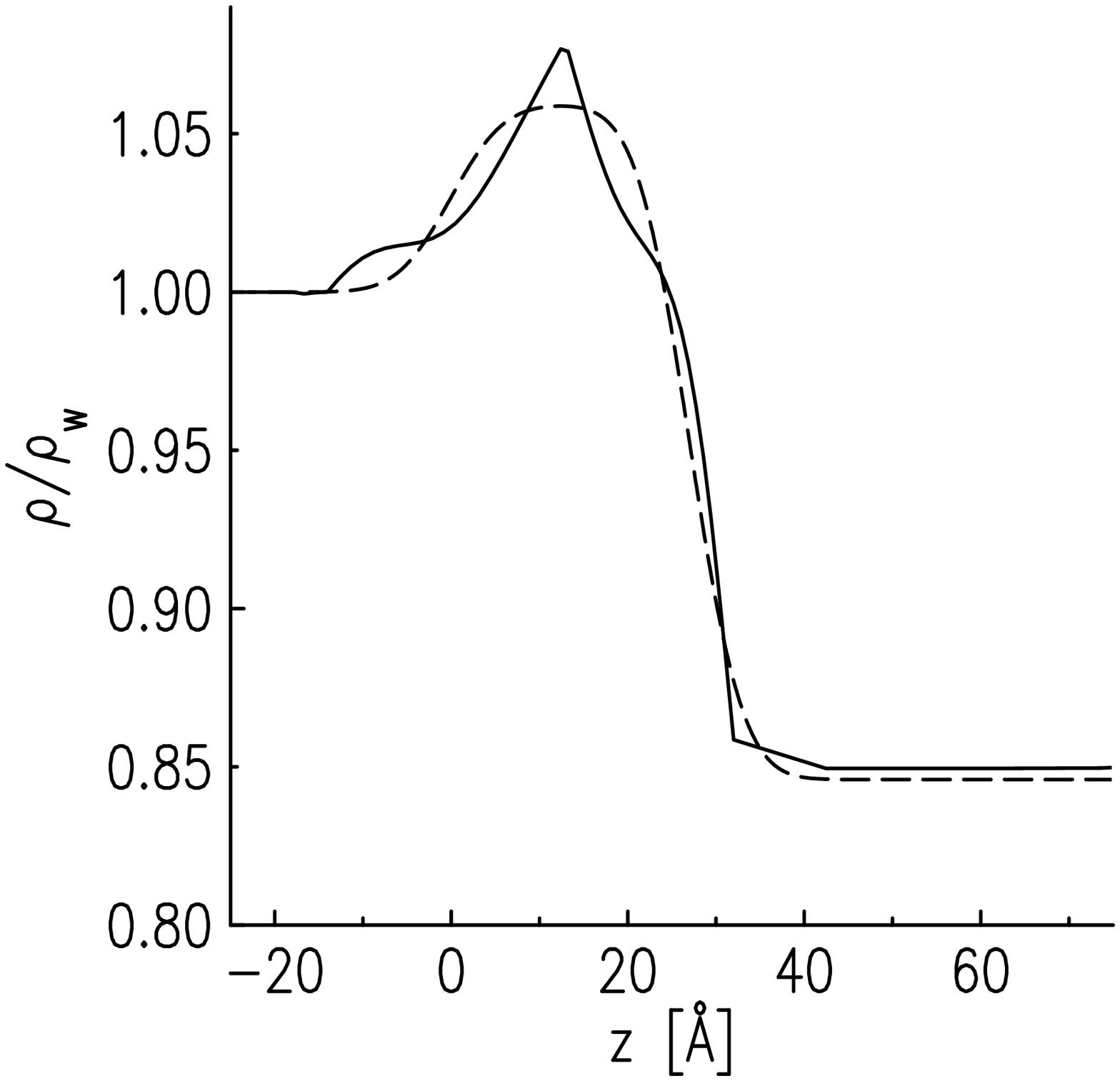, width=0.45\textwidth}

\small {\bf Figure 5.} \it Reconstructed electron density profiles $\rho(z)$ for a monolayer adsorption octadecanamide film at the toluene -- water interface for pH$\approx$11, normalized to the electron density in water under standard conditions ( $\rho_w=0.333$ {\it e$^-$/}{\AA}$^3$): solid and dashed curves correspond to model-independent and model (Eq. (1)) calculations. The position of the boundary between the polar region of molecules in the Gibbs monolayer and water is set at $z = 0$.

\end{figure}

The disregard of absorption in the medium in this approach imposes a limitation on the thickness
$L\ll\lambda^2q_z^{max}/4\pi\delta(z)$ of structure being reconstructed \cite{17}. Since the value of polarizability for water at wavelength $\lambda=0.825$\,\AA{}. $\delta \sim 2 \times 10^{-6}$, the admissible thickness of the reconstruction region ($L\gg 1000$\,\AA) is quite sufficient for determining the structure of the adsorption layer correctly \cite{46}.

\begin{figure}
\hspace{0.15in}
\epsfig{file=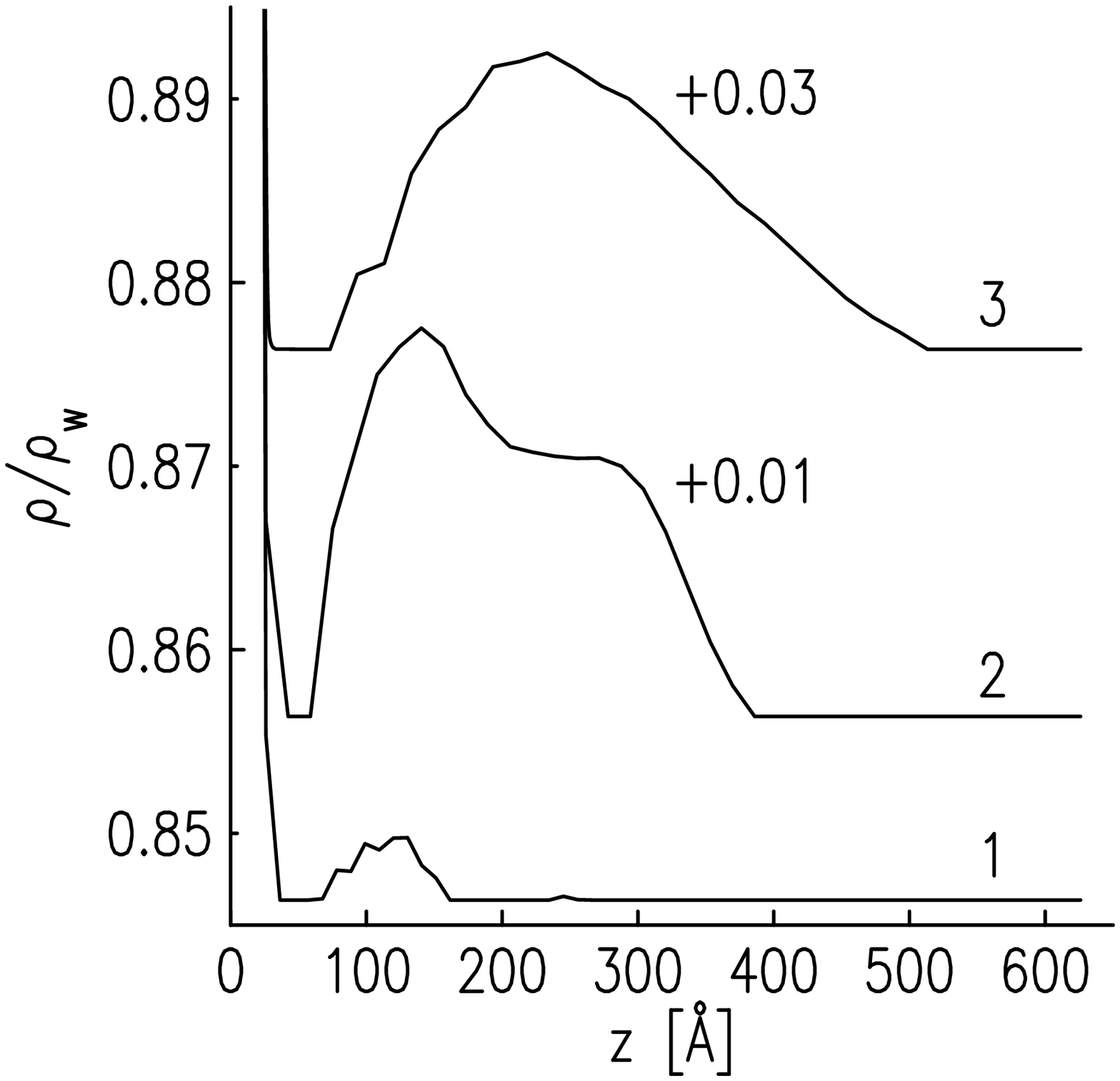, width=0.45\textwidth}

\small {\bf Figure 6.} \it Reconstructed electron density profiles $\rho(z)$ for the monolayer adsorption octadecanamide film at the toluene -- water interface normalized to the electron density in water under standard conditions ($\rho_w=0.333$ {\it e$^-$/}{\AA}$^3$): (1) pH$\approx 11$; (2) pH$\approx 7$; (3) pH$\approx 2$. For convenience of comparison, profiles 2 and 3 are shifted along the ordinate axis by $+0.01$ and by $+0.03$, respectively. The position of the boundary between the polar region of molecules in the Gibbs monolayer and water is set at $z = 0$.

\end{figure}

\vspace{0.25in}
{\bf RESULTS AND DISCUSSION}

The solid and dashed curves in Fig. 5 show electron density profiles $\rho(z)$ along the normal to the toluene -- water interface for pH$\approx 11$, which were reconstructed using the model-independent and model approaches, respectively. These curves correspond to a crystalline monolayer adsorption film with $A=18\pm1$\,{\AA}$^2$. Figure 6 shows reconstructed electron density profiles $\rho(z)$ for all systems in a wider range of $z$ as compared to that in Fig. 5. It demonstrates a qualitative change in the interface structure upon a decrease in the pH level from 11 to 2, which is due to the formation of a broad layer in the $z$ interval from $z^\prime \sim 70$\,{\AA} to $z_2\sim 500$\,{\AA} (see Fig. 1). With decreasing pH level, thickness $z_1$ and area $A$ per molecule in the Gibbs monolayer (for $0<z<z_1$) almost remain unchanged. It should also be noted that there is a sharp boundary between the monolayer and the broad layer in the form of region $z_1< z^\prime <70$\,{\AA} with an electron density of about $\sim \rho_t$. At the same time, for $z \sim z_2$, the boundary between the condensed micelle layer and the volume of the aromatic solvent is rather more diffuse than sharp.

As a rule, experiments with dissolvable adsorption layers are interpreted in terms of Gibbs adsorption \cite{47,48,49,50}. The applicability of such an approach is limited to systems with a low concentration of diphylic substance in a nonpolar organic solvent (true solution). At a concentration exceeding a certain critical concentration of amphiphilic substance, micelles that are in thermodynamic equilibrium with monomers as well as with the surface are formed in the solution \cite{51, 52}. It should also be noted that a microemulsion (three-component phase of a substance) could be formed in the Gibbs triad "nonpolar hydrocarbon solvent -- surfactant -- aqueous solution of electrolyte"{} \cite{53, 54}. In our experiment, volume concentration  $c \approx 5$\,mmol/L of octadecanamide in toluene was substantially higher than the typical critical concentration (lower than $<1$\,mmol/L) of micelle formations for nonionogenic surfactants in the aromatic hydrocarbon, but was not high enough for emulsification of the system \cite{55}. We attribute the formation of a thick layer upon a decrease in pH of the water subphase to the deposition of octadecanamide micelles in the field of the electric double layer at the toluene -- water interface (electrolyte solution).

In a nonpolar organic solvent (toluene), the micelle core is formed by hydrophylic polar groups -ÑÎNÍ$_2$, while hydrophobic tails of octadecanamide -C$_{17}$H$_{35}$ form the outer shell (inverted micelle). The minimal radius of a spherical micelle is $\sim L \approx 26$\,{\AA}. If we assume that the packing density of octadecanamide molecules in micelles and in the Gibbs monolayer is
approximately the same, the surfactant concentration in the micellar layer for pH $\approx 2$ is smaller than in the
model Gibbs layer by $(\rho_1-\rho_t)/(\rho_2-\rho_t)\sim 4 \div 5 $ times. At the same time, the broad layer is thicker than the
monolayer by $(z_2-z_1)/z_1 \sim 15 \div 20$ times. Therefore, the amount of octadecanamide in the micellar layer is
larger than its amount in the monolayer by 3 -- 4 times. Consequently, the surface concentration of nonionogenic surfactant at the oil -- water interface is determined by the pH level of the aqueous subphase, i.e., by the direction of polarization in the electric double layer at the boundary of the water phase or by the sign of the electric potential at the monolayer -- electrolyte solution interface.

The shape of the reconstructed electron density profiles $\rho(z)$ also indicates the existence of the plane of
the closest approach of micelles to distance $z^\prime \sim 70$\,{\AA} from the interface, and the emergence of the depletion
region in the interval $z_1< z^\prime < 70$\,{\AA} is a quite unexpected feature of the profile. Such an interface polarization
pattern is probably a manifestation of specific electrostatic and steric effects in the interaction of
micelles with the surface and requires a detailed investigation that is beyond the scope of this article. It
should only be noted that the existence of the closest approach plane for micelles has not been discussed in
earlier publications, because it is difficult to reliably establish its presence using the model approach to the
reconstruction of the $\rho(z)$ profile \cite{9, 14, 16}. This is primarily because with such an approach, the ambiguity
in determining these parameters rapidly increases as a rule upon an increase in the number of model
structure parameters.

Thus, the structure of the dissolvable adsorption octadecanamide film at the toluene -- water interface strongly depends on the pH level of the water subphase. At a high pH level, the film has the state of a solid monolayer of a thickness of approximately $26$\,{\AA}, in which aliphatic octadecanamide tails are extended along the normal to the surface. At a low pH level, the thickness of the surface structure increases to approximately $\sim 500$\,{\AA} due to condensation of surfactant micelles; as a result, the amount of the adsorbed substance increases by 3 -- 4 times (i.e., significantly). According to our results, the structure of the Gibbs monolayer in this case is independent of the pH level of the subphase. Using the model-independent approach, we could establish the existence of the plane of the closest approach of surfactant micelles to the interface, which lies at a distance of about $\sim 70$\,{\AA} from it in the pH interval from 2 to 11. In our opinion, the effect of micelle condensation at the interface upon a decrease in the pH level, which is demonstrated in this
study, is caused by the change in the direction of polarization in the electric double layer at the water phase boundary.

\vspace{0.25in}
{\bf ACKNOWLEDGMENTS}

The work at the National Synchrotron Light Source was supported by the US Department of Energy (contract no. DE-AC02-98CH10886). The work at the X19C beamline was supported by the ChemMatCARS National Synchrotron Resource, University of Chicago, University of Illinois at Chicago, and Stony Brook University.The theoretical part of this work was supported by the Russian Science Foundation (project no. 18-12-00108).

\end{document}